\documentclass[10 pt]{article}
\usepackage{epsfig}
\begin{document}

\title{\bf Quantum Interference: an experimental proposal, and possible Engineering applications.}

\author{A.Y. Shiekh\footnote{\rm shiekh@dinecollege.edu} \\
             {\it Din\'{e} College, Tsaile, Arizona, U.S.A.}}

\date{}

\maketitle

\abstract{A proposal for an experiment to look at some possibly novel aspects of quantum interference is presented, along with some Engineering applications that might result.}

\baselineskip .7 cm

\section{Introduction}
In a series of papers \cite{Shiekh1, Shiekh2, Shiekh3, Shiekh4} the author has tried to take advantage of some potentially neglected aspects of quantum interference, but the details of how that interference might be achieved were not correctly explained; in the author's former works the two beams were brought into near parallel overlap in free space, and this turned out not to be sufficient\footnote{private communication with Professor Philippe Eberhard}, while here that has been remedied by a semi-silvered mirror that brings the two beams into more perfect overlap.

It should perhaps also be emphasized that entanglement is not being used in this proposal, in keeping with the no-communication theory \cite{Ghirardi1, Ghirardi2, GerjuoySessler}, but rather the self interference of a single particle is being employed, to both augment the ability of a quantum computer and possibly achieve paradox free superluminal communication.

\section{Quantum Interference}
\begin{figure}[here] 
   \centering
   \includegraphics[width=4in]{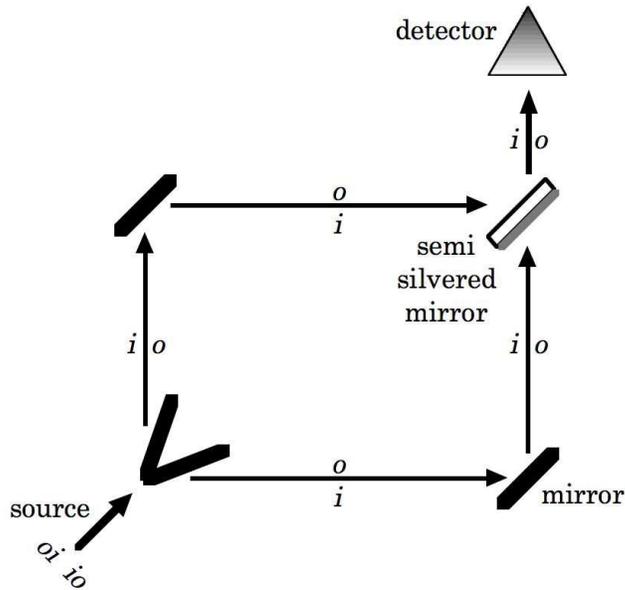} 
   \caption{Doubling Interferometer}
   \label{fig:doubling}
\end{figure}

Suggested here is a proposed configuration\footnote{the author is not competent to comment on the practical difficulties that might be encountered in the implementation of this proposal} for quantum interference based on a variation-on-a-theme of the Mach-Zehnder interferometer; the difference being that the particle's wave-function is cut into left and right hand parts by a mirror slicer, and the two pieces are then overlapped using the usual beam splitter. Interference is arranged such that the two halves align correctly when they overlap; $i$ and $o$ in Figure \ref{fig:doubling} refer to the inner and outer parts of each half.

For the horizontal output, the lower path undergoes three $\pi$ phase changes, while the upper path undergoes two, putting them into destructive interference. For the vertical output, the lower path undergoes two $\pi$ phase changes, while the upper path also experiences two, putting them into constructive interference. What has been achieved overall is a wave-function of half the volume and twice the amplitude of the original, which would seem to defy unitarity conservation, since for the original, normalized, wave-function
\begin{equation}
\int |\Psi|^2 dV = 1
\end{equation}
while now one has
\begin{equation}
\int |2 \Psi|^2 \frac{dV}{2}
\end{equation}
which yields a doubled normalization.

If such an effect is actually permitted by Nature, rather than a violation of unitarity, one might anticipate this wave-function amplification to be automatically globally re-normalized, a little like what happens upon the act of measurement. There is no conflict  with the unitarity conservation built into Schr\"odinger's equation, as it is the cutting and recombination, not the evolution, that is responsible for the effect being looked at here; despite this, there is an attempt to show that the configuration can be viewed as multi-valued like the Aharonov-Bohm case \cite{AharonovBohm}, and that unitarity actually is preserved on the resulting Riemann space \cite{Shiekh4}.

It has been commented\footnote{private communication with Professors Philippe Eberhard and Giancarlo Ghirardi} that this anticipated global renormalization is an extension of traditional quantum theory, although one that in the view of the author is necessary in light of the above observations. The question that now poses itself is what this extension might imply, and if any contraction ensues.

\subsection{Quantum Computing}
The renormalization effect of above might be used to amplify solutions in a quantum computer. 

\begin{figure}[here] 
   \centering
   \includegraphics[width=4in]{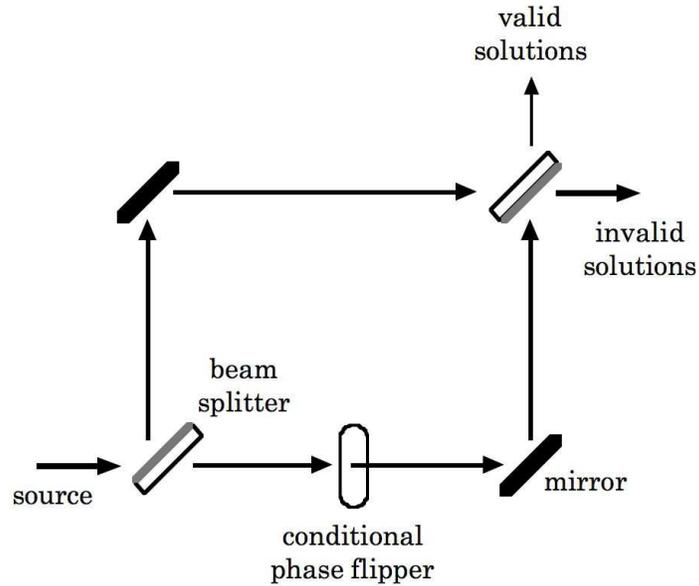} 
   \caption{Mach-Zehnder separation}
   \label{fig:MachZehnder}
\end{figure}

For the purposes of illustration, start with the following three qubit Hadamard state (leaving out normalizations for clarity)
\begin{eqnarray}
\left| \psi  \right> &=&
(\left| 0 \right> + \left| 1 \right>)
(\left| 0 \right> + \left| 1 \right>)
(\left| 0 \right> + \left| 1 \right>) \\
&=&
 \left| 0 0 0 \right> + 
 \left| 0 0 1 \right> +
 \left| 0 1 0 \right> +
 \left| 0 1 1 \right> + \ldots +
 \left| 1 1 1 \right>
\end{eqnarray}
and apply the decision function to these exhaustive candidate solutions to mark valid solutions alone by inverting their phase in one arm of a Mach-Zehnder interferometer\footnote{in practice the calculation would probably be done outside of the interferometer, and only the phase flip done within it} shown in Figure \ref{fig:MachZehnder}. 

The valid solutions will then appear in the vertical output, while the invalid ones will be in the horizontal output, since the vertical output will carry
\begin{equation}
\begin{array}{c}
-        \left| 0 0 0 \right> 
+ {\bf \left| 0 0 1 \right>} 
-        \left| 0 1 0 \right> 
+ {\bf \left| 0 1 1 \right>} - \dots 
-        \left| 1 1 1 \right> 
\\
+       \left| 0 0 0 \right> 
+ {\bf \left| 0 0 1 \right>} 
+       \left| 0 1 0 \right> 
+ {\bf \left| 0 1 1 \right>} + \dots 
+        \left| 1 1 1 \right> 
\end{array} 
\end{equation}
to expose the desired valid solutions
\begin{equation}
{\bf \left| 0 0 1 \right>} +
{\bf \left| 0 1 1 \right>}
\end{equation}
having assumed, for the sake of argument that solutions $001$ and $011$ satisfy the function;
while the horizontal output will have
\begin{equation}
\begin{array}{c}
+ \left| 0 0 0 \right>
- {\bf \left| 0 0 1 \right>}
+ \left| 0 1 0 \right>
- {\bf \left| 0 1 1 \right>} + \dots
+ \left| 1 1 1 \right> \nonumber 
\\
+       \left| 0 0 0 \right> 
+ {\bf \left| 0 0 1 \right>} 
+       \left| 0 1 0 \right> 
+ {\bf \left| 0 1 1 \right>} + \dots 
+        \left| 1 1 1 \right> 
\end{array} 
\end{equation}
to expose the many more invalid solutions.
\begin{equation}
{\left| 0 0 0 \right>} +
{\left| 0 1 0 \right>} + \dots
+  \left| 1 1 1 \right>
\end{equation}

This physical separation of the good and bad solutions is not in itself sufficient to identify the valid solutions, since they are in general overpowered by the much greater number of invalid solutions; however, after this separation one can repeatedly apply the doubling interferometer introduced above to the output arm containing the valid solutions, to achieve exponential amplification.

\subsection{Faster-than-light communication}
\begin{figure}[here] 
\centering
\includegraphics[width=4in]{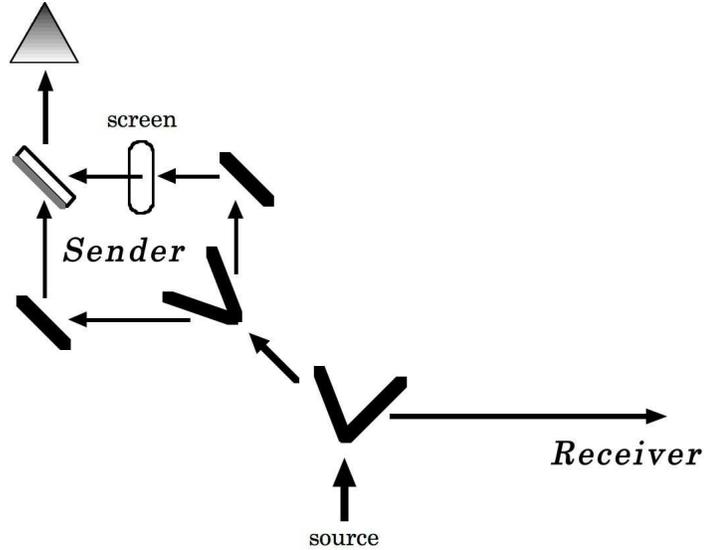} 
\caption{Faster-than-light transmission}
\label{fig:FTL}
\end{figure}

The suggested renormalization effect, if it really occurs, also seems to imply the ability to fabricate a faster-than-light device \cite{Fayngold}, see Figure \ref{fig:FTL},
as one could then selectively put a screen in the path to allow or not the renormalization to take place; an effect that would be seen at the receivers end as a fluctuating beam intensity. So a faster than light transmitter of information (but not energy or matter) might be possible.

\begin{figure}[here] 
   \centering
   \includegraphics[width=3in]{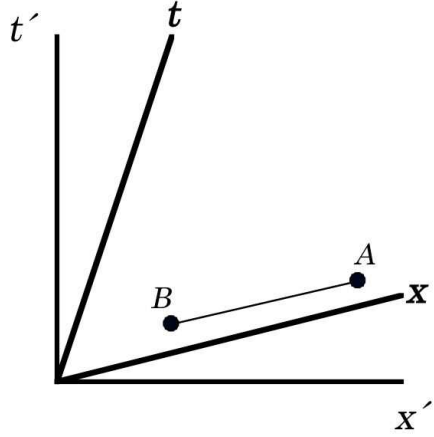} 
   \caption{Space-Time diagram}
   \label{fig:SpaceTime}
\end{figure}

The above proposal would seem more plausible if it can be demonstrated that no paradox arises from its supposed ability to communicate ÔinstantlyÕ over indefinite distances; namely, that no use can be made of a communication to alter events in the past. It is said that the collapse of the wave-function happens `instantly', but as is well known, relativity does not respect this concept; what is instant in one frame is not in another. It also does not seem reasonable that a moving measuring device would instigate a different collapse from a stationary one, and one way around this dilemma is that there is a preferred frame \cite{EberhardRoss, Eberhard} (a quantum aether) in which the collapse occurs. Up until now this was not a pressing issue for, although it would alter the cause and effect ordering for the measuring of an EPR pair, the end result was not influenced by which end made the measurement first. This uneasy state of affairs is brought to a head here, but fortunately the above suggestion that the collapse occurs in some preferred frame also severs to prevent the faster than light proposal from being able to communicate into the past. Figure \ref{fig:SpaceTime} shows the instantaneous collapse in the preferred ($x$, $t$) space-time frame, and while it could be used to send information back in time (from the perspective of $x'$, $t'$), it could not be used to alter the past, as the return signal would move forward in time by at least an equal amount,
thus resolving any potential paradox.

Faster than light communication also has a bearing on computation, as present day computing speeds of 4 GHz imply a light travel distance of only 7.5 cm; so the two offsprings of this work are not unrelated.

\section{Conclusion}

Faster than light communication may be possible using the amplification mechanism proposed above, and without any paradoxical powers accompanying the device.

The same amplification mechanism might also be put to work in augmenting the power of a quantum computer, and at the same time use only the digital aspects of quantum theory for calculation, which might greatly simplify the error correction aspects \cite{NielsenChuang}. The problem with analogue systems is that errors, all be them small, creep into {\em all} aspects of the system in any finite time, and so are impossible to remove completely. In contrast, digital systems have the advantage that the probability of an error is generally small, albeit that the error itself, if seen, is large; so for a small time interval, it is very unlikely that all aspects of the system find themselves in error.

\end{document}